\documentclass[aps,prb,twocolumn,showpacs,superscriptaddress]{revtex4}
\usepackage{graphicx,color}% Include figure files
\usepackage{amsmath}
\usepackage{amssymb}
\usepackage{textcomp}
\usepackage{bm}

\renewcommand{\Re}{\mathop{\text{Re}}\nolimits}

\newcommand{\ket}[1]{|{#1}\rangle}

\newcommand{\sinc}{\mathop{\text{sinc}}\nolimits}

\definecolor{dgreen}{rgb}{0,0.5,0}

\definecolor{delete}{cmyk}{0.5,0,0,0}

\begin{document}
%\preprint{}

\title{Probing Cooper pairs with Franson interferometry}



\author{Vittorio Giovannetti}
\affiliation{NEST-CNR-INFM \& Scuola Normale Superiore, Piazza dei Cavalieri 7, I-56126 Pisa, Italy}
\author{Kazuya Yuasa}
\affiliation{Department of Physics, Waseda University, Tokyo 169-8555, Japan}

%\thanks{}


%\date{\today}
\date[]{September 12, 2012}

\begin{abstract}
A setup based on the Franson optical interferometer is analyzed, which allows us to detect the coherence properties of Cooper pairs emerging via tunneling from a superconductor in contact with two one-dimensional channels. 
By tuning the system parameters we show that both the internal coherence of the emitted Cooper pairs, which is proportional to Pippard's length, and  the de Broglie wavelength of their center-of-mass motion can be measured via current-current correlation measurements. 
\end{abstract}

%\keywords{}

\maketitle

\section{Introduction} 
Nonlocal interferometry is a highly successful tool for characterizing the nonclassical properties of  quantum correlated systems.\cite{ref:FransonReview2,TITTEL}
In particular, the Franson interferometer \cite{ref:FransonInt}
 suits for studying the temporal and spatial coherence of  pairs of correlated particles emitted by a quantum source. 
It consists in recording time-resolved  
coincidence events at selected output ports of a couple of Mach-Zehnder interferometers (MZIs), each fed by one of the particles of each pair, whose delays 
are properly set in order to discard 
the first-order coherence contributions (i.e., contributions which might arise from the interference between different single-particle paths).
In other words, the Franson setting is specifically designed to detect the global properties of the pairs (say, their de Broglie wavelength \cite{ref:Jacobson-PRL1995,ref:Barnett1998,ref:Fonseca-PRL1999,ref:Edamatsu-PRL2002}), 
treating them as unique objects which, independent of the spatial/temporal distribution of their constituents,  
can be delocalized in space and/or time.

Numerous realizations of the Franson scheme have been implemented in optics
probing  either two-photon states produced via a parametric down-conversion process,\cite{EXP,PhysRevLett.65.321,PhysRevLett.66.1142,PhysRevA.44.4552,PhysRevA.45.2052,ref:Shih-PRA1993,PhysRevA.47.R2472,PhysRevA.77.021801} or entangled photon-hole
states \cite{PhysRevLett.96.090402} produced via two-photon absorbing processes.
The same technique has also been experimentally tested  with pairs of surface plasmons excited with a two-photon entangled source,\cite{ref:Plasmons} and it has been proposed  for characterizing coherent transport in two-dimensional mesoscopic conductors.\cite{PhysRevLett.92.167901,PhysRevLett.103.076804,ref:LucaVittorioSaro-TimeBin} 
Interferometric setups required for such proposed experiments for electrons in solids have already been realized (e.g., see Refs.~\onlinecite{ref:Antibunching-Ji,ref:MachZenderElectron-PRL100,ref:MachZenderElectron-PRB78,ref:MachZenderElectron-PRB79}), 
and several correlation experiments have been carried out,\cite{ref:Antibunching-Yamamoto,ref:Antibunching-Oliver,ref:Antibunching-Henny,ref:Antibunching-Neder,ref:BuettikerReview2000,ref:Samuelsson-HBT,ref:SamuelssonButtikerReview}
also with superconducting sources of electrons.\cite{ref:CooperSplitterNature,ref:CooperSplitterPRL}

Here we discuss the possibility of  employing a Franson interferometer to study
 Cooper pairs emitted from a superconductor attached to a multi-port semiconductor device. 
 Specifically the aim of our work  is to show how
 the interference fringes measured at the outputs of the device can be used to characterize the two-particle wave function of the emitted Cooper pair.
Previous works have proposed to detect  the entanglement present in the spin degrees of freedom of a Cooper pair via correlation measurements.\cite{ref:BuettikerReview2000,PhysRevB.61.R16303,PhysRevB.65.075317,PhysRevLett.88.197001,PhysRevB.66.161320,PhysRevB.69.125326,PhysRevB.70.115330,ref:BurkardReview-JPC2007,ref:CooperBunchRapid,ref:CooperBunch,ref:CooperInt}
The Franson interferometer we consider here, on the other hand, is suited for probing the coherence properties of the \textit{spatial} degrees of freedom 
of the emitted Cooper pair.
By looking at the interference fringes in the coincidence counts at the output ports, varying the relevant parameters of the Franson interferometer, we can measure the pair correlation length of the emitted Cooper pair, which is proportional to Pippard's length characterizing the extension of the Cooper pair in the superconducting source, as well as the momentum of the center-of-mass motion of the emitted Cooper pair, i.e., the de Broglie wavelength of the emitted Cooper pair as a single object.

This paper is organized as follows. In Sec.\ \ref{sec:one} we recall some basic facts on the Franson interferometer enlightening how it allows one to 
characterize two-particle quantum coherence in the optical case for several parameter configurations. 
Then in Sec.\ \ref{sec:two} we discuss how it could be applied for the study of the Cooper-pair wave function. 
The paper ends with Sec.~\ref{Conclusions} with conclusions and remarks.

%%%%%%%%%%%%%%%%%%%%%%%%%%%%%%%%%
\begin{figure}[b]
\includegraphics[width=0.48\textwidth]{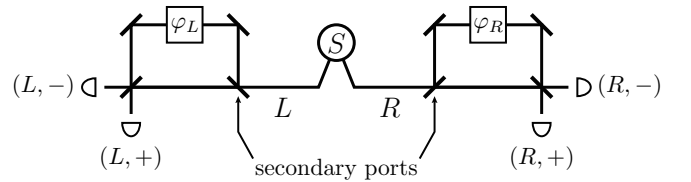}
\caption{Schematics of the Franson interferometer (optical implementation). The EPR-like state (\ref{input}) is injected by the source $S$ along the two optical
paths $L$ and $R$, while the coincidence counts are recorded at the detectors, e.g., $(L,-)$ and $(R,-)$, to reveal the two-photon coherence of the emitted bi-photon state.}
\label{fig:Setup}
\end{figure}
%%%%%%%%%%%%%%%%%%%%%%%%%%%%%%%%%
\section{The Franson interferometer} \label{sec:one}
The aim of the Franson interferometric scheme is to reveal the coherence properties of the field emitted by a source $S$ which produces particles in correlated pairs,
 by recording coincidence events that take place at distant detectors.\cite{ref:FransonReview2,ref:FransonInt}
An example of the setup is sketched in  Fig.\ \ref{fig:Setup}, where $S$ is a nonclassical source which emits two
entangled EPR-like photons that propagate towards the two MZIs placed along the left and right arms of the figure.
Specifically  one photon is injected along the  optical path  ``$L$'' with longitudinal wave vector $k_0+k$
 and polarization $\sigma$, 
while the other is injected in the path ``$R$'' with wave vector $k_0-k$
 and the opposite polarization $\bar{\sigma}$.  
 Indicating the electromagnetic vacuum state with $\ket{\O}$, the input of the interferometer  can be expressed~as 
 \begin{equation}\label{approxstate}
|\Psi\rangle \simeq |\O\rangle + |\Psi^{(2)}\rangle + \cdots, 
\end{equation} 
 with $|\Psi^{(2)}\rangle$ being the two-photon state 
\begin{equation}
\ket{\Psi^{(2)}}
:=
\sum_{\sigma}
 \int dk \,\Psi_k b_{k_0+k,\sigma}^{(L)\dag}b_{k_0-k,\bar{\sigma}}^{(R)\dag}\ket{\O},\label{input}
\end{equation} 
where, assigning a positive momentum for a photon out-going from the source $S$, $b_{k,\sigma}^{(L/R)}$ and $b_{k,\sigma}^{(L/R)\dag}$ are the (bosonic) annihilation and creation operators, respectively, associated with the spatial direction $L/R$ of the setup. 
According to  (\ref{input}) the two photons of $|\Psi^{(2)}\rangle$ individually propagate with average wave vector $k_0$, yielding a joint momentum  
$p_\text{pair}:={2\hbar k_0}$ 
of the pair (the de Broglie momentum of the system \cite{ref:Jacobson-PRL1995,ref:Barnett1998,ref:Fonseca-PRL1999,ref:Edamatsu-PRL2002}).
Vice versa their relative motion is characterized by a two-particle wave function $\Psi_k$,
centered at $k=0$ with width $\Delta k$, and normalized to half the mode spacing $\delta k$ of the optical path, times the amplitude probability associated with the emitting event.
In the space representation, this corresponds to having two correlated single-photon pulses with their mean positions delocalized along the line, and 
 with a relative distance which instead is coherently spreading according to the  amplitude distribution 
\begin{equation}
{\Phi}(x): =\int\frac{dk}{\sqrt{2\pi}}\Psi_k e^{ikx},
\label{eqn:PhiPsi}
\end{equation}
centered at $x=0$ and spread over an interval $\Delta x\sim 1/\Delta k$ (the first-order coherence length of the emitted field~\cite{EXP,PhysRevLett.65.321,PhysRevLett.66.1142,PhysRevA.44.4552,PhysRevA.45.2052,ref:Shih-PRA1993,PhysRevA.47.R2472,PhysRevA.77.021801}).

The state $\ket{\Psi}$ in (\ref{approxstate}) 
 can be seen as the state produced via two-photon emission 
by a three-level atom,\cite{ref:FransonInt} in which the decay of a high-energy level $\omega_0$ of long lifetime yields a couple of correlated pulses having opposite polarizations and complementary energies and momenta. 
Alternatively $|\Psi\rangle$ can be identified with the two-mode squeezed state  
 produced via a parametric down-conversion process through a weak but stable laser of frequency $\omega_0=2  ck_0$, 
 which is pumped on a short $\chi^{(2)}$ nonlinear crystal under proper  phase-matching conditions.
In this case, in fact, defining the two-mode squeezing operator of the process 
\begin{equation}
\Omega:=  \label{squeezingoperator}
\exp\!\left[
\sum_\sigma\int dk\,(\Psi_k b_{k_0+k,\sigma}^{(L)\dag} b_{k_0-k,\bar{\sigma}}^{(R)\dag}
- \text{H.c.})
\right],
\end{equation}
with $\Psi_k$ being a linear function of the pump intensity,  
 the nonequilibrium steady  state (NESS) of the radiation emerging from the crystal and propagating along the optical paths $L$ and $R$ 
reads  $\Omega |\O\rangle$, which reduces to (\ref{approxstate})  when neglecting the higher-order  terms in the squeezing parameter $\Psi_k$.
For such an implementation   the delocalization of the two-photon mean position corresponds to the coherence length of the pump (up to a few meters),
while the typical values are $\lambda_0=2\pi/k_0 \simeq 700 \,\text{nm}$  and $\Delta x\simeq 500\, \mu\text{m}$, \cite{EXP,PhysRevLett.65.321,PhysRevLett.66.1142,PhysRevA.44.4552,PhysRevA.45.2052,ref:Shih-PRA1993,PhysRevA.47.R2472,PhysRevA.77.021801,ref:KwiatPDCEntanglement,ref:Grice-PRA1997}
respectively.

\subsection{Counting coincidences}
\label{sec:Coincidences}
After propagating along its corresponding interferometric arm, each photon 
is sent through its corresponding unbalanced MZI
and is detected (without discriminating  its polarization) at one of the associated  output ports, where 
coincidences are recorded, say between detectors $(L,-)$ and $(R,-)$ of Fig.\ \ref{fig:Setup} (in the following we focus on the coincidences between these ports without loss of generality).
The statistics of such events can be expressed in terms of the 
infinitesimal  joint probability of one photon being revealed by detector $(L,-)$  at time $t_L=t$ within $dt$ 
 and the other by $(R,-)$ at time $t_R=t+\tau$ within $d\tau$, i.e., 
\begin{equation} \label{def:P2}
d P^{(2)} = \eta^2I^{(2)}(x_L,t;x_R,t+\tau)\, d t  \,d \tau,
\end{equation}
where $\eta$ is the effective quantum efficiency of the detectors and $I^{(2)}(x_L,t;x_R,t+\tau)=I^{(2)}(\tau)$ 
is the fourth-order correlation function, 
\begin{align}
I^{(2)}(\tau)
:={}&
\sum_{\sigma_L,\sigma_R}
\langle
\psi
_{\sigma_L}
^{(L,-)\dag}
\psi_{\sigma_R}
^{(R,-)\dag}
\psi
_{\sigma_R}
^{(R,-)}
\psi_{\sigma_L}
^{(L,-)}
\rangle 
\nonumber\\
={}&\frac{1}{16\pi}|2{\Phi}(c\tau) \cos\phi 
-e^{i\varphi} {\Phi}(c \tau+\delta L)
\nonumber\\
&\qquad\qquad\qquad\quad\ %
{}-e^{-i\varphi} {\Phi}(c \tau-\delta L) 
|^2. 
 \label{eqn:I} 
\end{align}
Here 
$\langle {\cdots} \rangle$ stands for the expectation value over the input state $|\Psi^{(2)}\rangle$ [or $|\Psi\rangle$ in (\ref{approxstate}), the result being identical at the lowest order in the squeezing expansion], while
 $\psi_\sigma^{(\ell,\pm)}$ is the field operator at  the $(\ell=L/R,\pm)$ detector.  Apart from an irrelevant contribution from the secondary input port of the associated MZI (see Fig.\ \ref{fig:Setup}), the latter  is
 \begin{equation}
 \psi_\sigma^{(\ell,\pm)}:=\frac{1}{2}[\psi_\sigma^{(\ell)}(x_\ell,t_\ell)\pm e^{i\varphi_\ell}\psi_\sigma^{(\ell)}(x_\ell+\delta L_\ell,t_\ell)],
\label{freeout}
\end{equation}
where $x_\ell$ and $x_\ell + \delta L_\ell$ are, respectively, the
distances of the detectors from $S$ measured along the short and long paths of the MZI, $\varphi_\ell$ is an extra relative phase 
the photon might accumulate while propagating through the MZI, and given  the frequency of the mode  $\omega_k= c |k|$,
\begin{equation}
\psi_\sigma^{(\ell)}(x,t):=
\int \frac{dk}{\sqrt{2\pi}}b_{k,\sigma}^{(\ell)}e^{i (kx-\omega_k t)}
 \label{free}
\end{equation}
 is the free field along the optical path $\ell$, which connects $S$ to the corresponding MZI\@.
Finally Eq.\ (\ref{eqn:I}) has been computed by setting 
 $x_L=x_R$, $\delta L_L=\delta L_R=\delta L$  (we assume this setting in the following), and
 introducing the parameters
 \begin{gather}
\phi:=k_0\delta L+(\varphi_L+\varphi_R)/2,\nonumber\\
 \varphi:= (\varphi_L-\varphi_R)/2.
\label{varphi}
\end{gather}
With these definitions the first term within the square modulus of ({\ref{eqn:I}) is identified 
 with  the interference between 
 (i) a pair of photons both going the long paths of the MZIs and (ii) both going the short paths [LL + SS interference; see Fig.\ \ref{fig:InterferingProcs}(a)].
The other two terms of  (\ref{eqn:I})
instead  represent the processes in which one of the two photons goes the long path while the other goes the short path [LS and SL; see Fig.\ \ref{fig:InterferingProcs}(b)].  
%%%%%%%%%%%%%%%%%%%%%%%%%%%%%%%%%
\begin{figure}
\begin{tabular}{l}
(a)\\[-3.5truemm]
\includegraphics[width=0.45\textwidth]{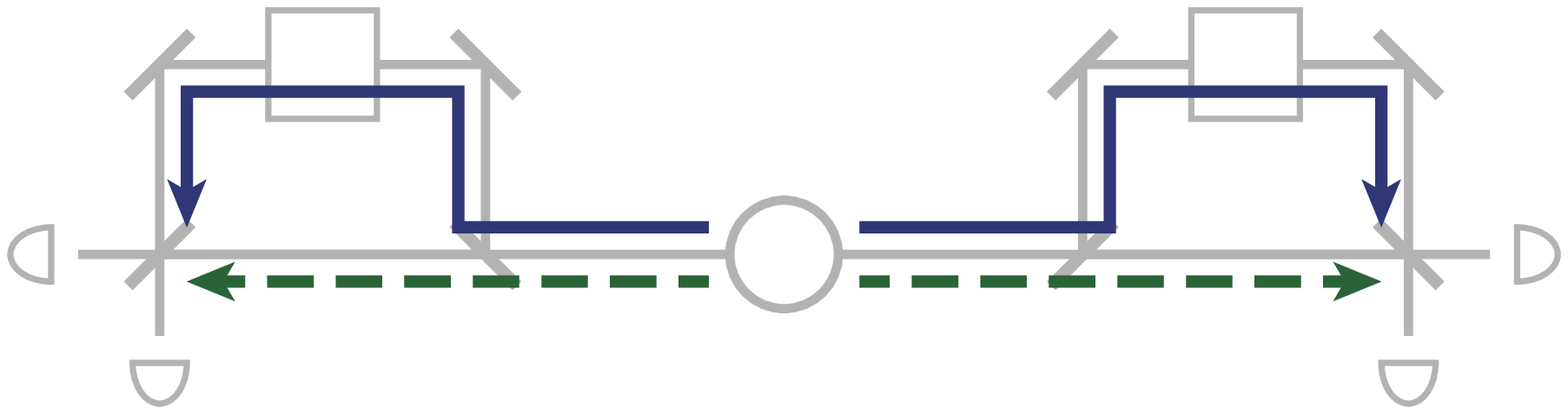}
\\
(b)\\[-3.5truemm]
\includegraphics[width=0.45\textwidth]{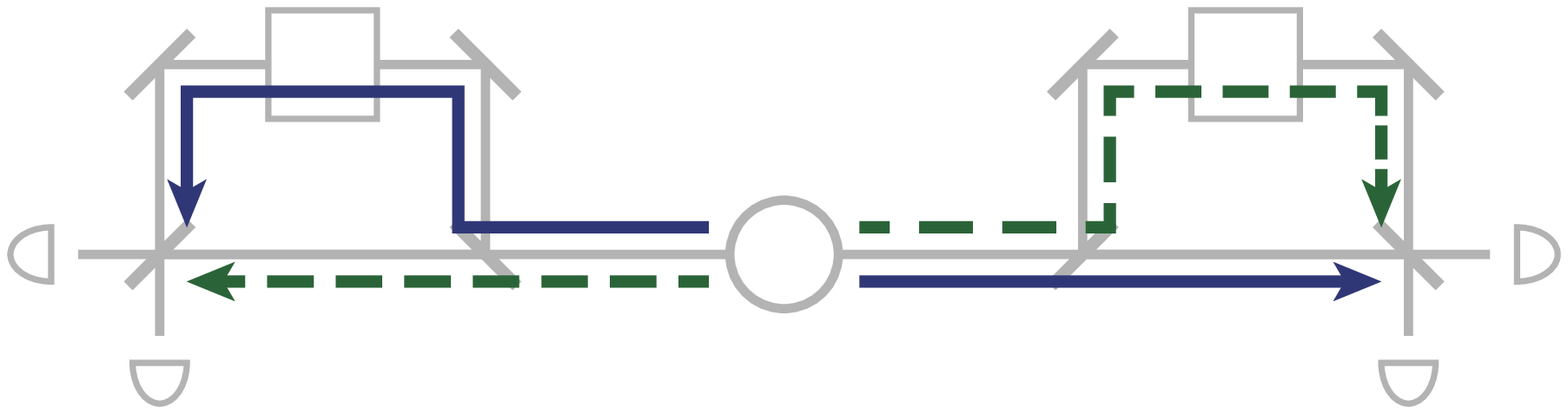}
\end{tabular}
\caption{(Color online) Four interfering processes:
(a) long-long (LL) (solid lines) and short-short (SS) (dashed lines), (b) long-short (LS) (solid lines) and short-long (SL) (dashed lines).
}
\label{fig:InterferingProcs}
\end{figure}
%%%%%%%%%%%%%%%%%%%%%%%%%%%%%%%%%

In a similar way we can also evaluate the statistics of the single-photon detection events
 at the two detectors $(L,-)$ and $(R,-)$ by means of the quantity
 \begin{equation} \label{def:P1}
d P^{(1)}_{L/R} = \eta I^{(1)}(x_{L/R},t)\, d t,
\end{equation}
which measures the infinitesimal probability of getting a detection at $(L/R,-)$ in the time interval $[t, t+dt]$
with
\begin{align}
&I^{(1)}(x_{L/R},t)
=\sum_\sigma
\langle
\psi_\sigma^{(L/R,-)\dag}
\psi_{\sigma}
^{(L/R,-)}
\rangle
\nonumber\\
&\quad
=\frac{1}{4\pi}\int dx\,|\Phi(x)-e^{i(\varphi_{L/R}+k_0 \delta L)}\Phi(x+\delta L)|^2.
\label{eqn:I1}
\end{align}

To clarify to which extent the interference recorded in (\ref{eqn:I}) provides a distinctive signature of the coherent behavior of the  two-photon ``compound" 
as a unique object, we need to look more closely at the  detection process. 
In particular we need to take account of the fact that in general the coincidences are registered over a finite time window $\Delta \tau$, whose extent might be larger than the typical time-scale of the system (a fact that might blur the two-photon features). 
Inserting (\ref{eqn:I}) into (\ref{def:P2}) and integrating the resulting expression  for $t\in[0,T]$ and for $\tau\in[0,\Delta \tau]$, 
we compute the 
average number of coincidence events 
\begin{equation}
N^{(2)}:=\int_{0}^T  \int_0^{\Delta \tau} d P^{(2)}, \label{NUMBER2} 
\end{equation}
where we assume that $T$ is the largest 
time scale of the system (still however  smaller than the coherence time of the pump that generates the bi-photon state). We  compare this with 
 the average number of single-detection events
 at the two detectors measured over  the time interval $T$, i.e.,
\begin{equation}
N^{(1)}_{L/R}:=  \int^{T}_0 d P^{(1)}_{L/R}.
\end{equation}
That is, the relevant quantity for our discussion is the coincidence ratio between the average number of 
 coincidences  vs the average number of couples  detected by the two detectors in $[0,T]$,
\begin{equation}
\mathcal{R}:=\frac{N^{(2)}}{N_L^{(1)}N_R^{(1)}}.
 \label{correlations}
\end{equation}

Under these  conditions we now recognize the existence  of several operative regimes characterized by the different 
ratios among the imbalance length $\delta L$, the first-order coherence length $\Delta x$
of the two-photon state, and the coincidence window length $c\Delta \tau$. 
To analyze them we first fix $\varphi_R=\varphi_L$.
The average number of single-detection events then reads
\begin{equation} \label{n1def} 
N_{L/R}^{(1)}
=\frac{\eta T}{4\pi}\int dx\,|\Phi(x)-e^{i \phi}\Phi(x+\delta L)|^2.
\end{equation}

\subsection{Fine resolution in time $\Delta \tau \ll \Delta x/c$}\label{sec:FineReso}
Let us start by considering the ideal scenario where $\Delta \tau$ is smaller  than the first-order coherence length, i.e.,  $\Delta \tau \ll \Delta x/c$.\cite{NOTE}
We then get a simplified expression for $N^{(2)}$,
\begin{equation}
N^{(2)}\simeq
\frac{\eta^2T\Delta\tau}{16\pi}|2\Phi(0)\cos\phi 
-\Phi(\delta L)-\Phi(-\delta L)|^2, 
\label{ratedef} 
\end{equation}
where again the first term 
within the square modulus  corresponds to the LL + SS interference while the other two to the LS and SL interference effects.

\subsubsection{Low imbalance regime $\delta L \ll \Delta x$}
Suppose now that the MZIs are operated in the  \textit{low imbalance regime}  $\delta L \ll \Delta x$. This implies  ${\Phi}(0) \simeq {\Phi}(\pm \delta L)$, so that 
 Eqs.\ (\ref{ratedef}) and (\ref{n1def}) are further reduced to
  \begin{gather} \label{low}
N^{(2)}\simeq\frac{\eta^2 T\Delta \tau}{4\pi}|\Phi(0)|^2(1- \cos\phi)^2, 
\\
N_{L/R}^{(1)}\simeq   \frac{\eta T}{2\pi} \mathcal{X}(0)
   (1 - \cos\phi), \label{N1low}
 \end{gather}
respectively,  where we have introduced the envelope function
  \begin{equation} \label{ENVELOP}
 \mathcal{X}(\delta L):= \int dx\, \Phi^*(x) \Phi(x+\delta L).
 \end{equation}
We notice that  $N^{(2)}$ exhibits oscillations in $\phi$  of visibility 1 and  period
    $2\pi$, which however can be accounted for by the independent  single-particle interference effects taking places at the two MZIs.  
 Indeed the coincidence ratio for the present case is given by
 \begin{equation}
 \mathcal{R}=\frac{\gamma\Delta \tau}{T},
 \label{eqn:R11}
 \end{equation}
which is constant, with
   $\gamma =\pi |\Phi(0)/\mathcal{X}(0)|^2$ being a pure number that depends only on the input state (\ref{input}).

\subsubsection{High imbalance regime $\delta L \gg \Delta x$}
The situation changes dramatically if instead the MZIs are operated in the \textit{high imbalance regime} $\delta L\gg \Delta x$.
In this case in fact,  one has $|{\Phi}(0)| \gg |{\Phi}(\pm \delta L)|$ so that the LS and SL contributions to $N^{(2)}$
can be neglected, while the LL + SS interference dominates the coincidence counts (\ref{ratedef}), yielding
\begin{equation}
N^{(2)} \simeq  \frac{\eta^2 T \Delta \tau}{4 \pi}
|{\Phi}(0)|^2 \cos^2\!\phi,
\label{high}
\end{equation}
which still exhibits oscillations in $\phi$ of visibility 1. However the  periodicity is now $\pi$, which is half  the
one it had in the low imbalance case, and this cannot be accounted for by the 
single-photon events. Indeed, since $\Phi^*(x)\Phi(x+\delta L)\simeq0$,
the average number  of single-photon  detections (\ref{n1def}) is constant equal to
\begin{equation} \label{N1const}
N_{L/R}^{(1)}\simeq\frac{\eta T}{2\pi}  \mathcal{X}(0),
 \end{equation}
and the ratio oscillates with period $\pi$, 
\begin{equation}
\mathcal{R}
  = \frac{\gamma\Delta \tau}{T}\cos^2\!\phi.
\label{eqn:R12}
\end{equation}
Remembering (\ref{varphi}) one notices that the oscillations in (\ref{high}) are related to the de Broglie momentum $p_\text{pair}={2\hbar k_0}$ of the photon pair. On the other hand
the fringes in (\ref{low}) are related to $\hbar k_0$ (i.e., to the average momentum of each single component of the pair). The difference in the periodicity of (\ref{high}) and (\ref{low}) hence  reflects
the fact that in the former case the oscillations arise when the two-photon compound 
interferes with itself as a single object propagating with momentum ${2\hbar k_0}$, while in the latter case they arise from the interference of a single component of the pair.\cite{NOTE1}

\subsection{Low resolution in time $\Delta \tau \gg \Delta x/c, \delta L/c$}
Consider now the case where  $\Delta \tau$ is the largest time scale of the system (i.e., larger  than $\Delta x/c$ and $\delta L/c$).
Clearly when this happens part of the information associated with (\ref{eqn:I}) is washed away. 
Indeed in this case
  Eq.~(\ref{ratedef}) gets replaced by 
\begin{align}
N^{(2)}\simeq\frac{\eta^2T }{8\pi c} 
[& \mathcal{X}(0)(1+2\cos^2\!\phi) 
\nonumber \\
&{}-4  \Re\mathcal{X}(\delta L)\cos\phi 
+  \Re\mathcal{X}(2\delta L)],
\label{ratedeflong}
\end{align}
where $\mathcal{X}(\delta L)$ is the envelope function defined in (\ref{ENVELOP}).

In the low imbalance regime $\delta L \ll \Delta x$, this gives
\begin{equation}
N^{(2)}\simeq
\frac{\eta^2T }{4\pi c}\mathcal{X}(0) 
 (1- \cos\phi)^2, \label{ratedeflong1}
\end{equation}
which as (\ref{low})  presents only single-particle oscillations, with $N_{L/R}^{(1)}$ being expressed as in (\ref{N1low}), and the coincidence ratio reads
\begin{equation}
\mathcal{R}
=\tilde{\gamma}=\pi/cT\mathcal{X}(0).
\label{eqn:R21}
\end{equation}
On the contrary
in the high imbalance regime $\delta L \gg \Delta x$,  we get
\begin{equation}
N^{(2)}\simeq
\frac{\eta^2T }{8\pi c}\mathcal{X}(0) 
 (1+2\cos^2\!\phi) \label{ratedeflong2},
\end{equation}
which as in (\ref{high}) exhibits oscillations of period $\pi$ but with a reduced visibility, with $N_{L/R}^{(1)}$ being expressed as in (\ref{N1const}).
The coincidence ratio in this case is given by
\begin{equation}
\mathcal{R}
=\frac{\tilde{\gamma}}{2} (1+2\cos^2\!\phi),
\label{ratedeflong2R}
\end{equation}
which oscillates with period $\pi$, but with a reduced visibility.

As extensively discussed in Refs.~\onlinecite{EXP,PhysRevLett.65.321,PhysRevLett.66.1142,PhysRevA.44.4552,PhysRevA.45.2052,ref:Shih-PRA1993,PhysRevA.47.R2472,PhysRevA.77.021801}
this prevents one from detecting the coherence of the superposition (\ref{input}) (indeed there exists classically correlated input states, in which
the coherence of $|\Psi^{(2)}\rangle$  is broken that yields the same signal). Still the $\pi$-periodicity  of the oscillations of (\ref{ratedeflong2}) bears a clear signature of the 
de Broglie momentum $p_\text{pair}$.

\subsection{Recovering the two-particle de Broglie wavelength in the low imbalance  regime $\delta L \ll \Delta x$}
\label{sec:random}
The analysis presented in the previous section shows that, independent of the relative width of $\Delta \tau$, the worst-case scenario for recovering the two-photon features of the input state is the low imbalance
regime $\delta L\ll\Delta x$. Under this circumstance in fact the oscillation of $N^{(2)}$ can be just interpreted as due to the single-particle interference effect recorded
in $N_{L/R}^{(1)}$. 
Enforcing the high imbalance condition $\delta L\gg\Delta x$, on the other hand, suppresses the single-particle interference effect while keeping  the two-photon one.

Unfortunately  for some of the applications we have in mind (see the next section), choosing 
$\delta L\gg \Delta x$ could  be a challenging constraint. 
A possible way out is to add some correlated  noise in the interferometer which suppresses the single-particle coherence effect, while preserving the two-photon coherence. 
For instance this can be done by 
keeping $\varphi_L+\varphi_R$ constant while randomizing $\varphi=(\varphi_L-\varphi_R)/2$.  By doing so
in fact $N_{L/R}^{(1)}$ loses the dependence upon $\phi$  and reduces to the constant value given in (\ref{N1const}), while 
 Eq.\ (\ref{eqn:I}) gets replaced by 
\begin{align}
I^{(2)}(\tau)=\frac{1}{16\pi}\,\Bigl(
4 | {\Phi}(c\tau)|^2 \cos^2\!\phi
&{}+   | {\Phi}(c \tau+\delta L)|^2
\nonumber\\
&{} + | {\Phi}(c \tau-\delta L)|^2\Bigr).
\end{align}
Consequently, even if $\delta L\ll\Delta x$, we get
\begin{gather}
N^{(2)}
\simeq\frac{\eta^2T\Delta\tau}{8\pi}| {\Phi}(0)|^2
(1+2\cos^2\!\phi),
\label{eqn:N2Randomized}\\
\mathcal{R}
\simeq\frac{\gamma\Delta\tau}{2T}
(1+2\cos^2\!\phi)
\label{eqn:R31}
\end{gather}
for $\Delta\tau\ll\Delta x/c$, while Eqs.\ (\ref{ratedeflong2}) and (\ref{ratedeflong2R}) are reproduced for $\Delta\tau\gg\Delta x/c$:
in both cases the coincidence ratio oscillates as a function of $\phi$ with period $\pi$ but with half visibility.

\section{Cooper-pair emission} \label{sec:two}
Having clarified the basic idea of the proposal we now show how it could be applied for characterizing the two-particle wave function of a Cooper pair emitted from a superconductor into 
external metallic contacts via a tunneling process.  
The fundamental observation 
here is that, within the validity of the self-consistent mean-field approximation,\cite{ref:Tinkham}  the BCS state $\ket{\text{BCS}}$, which describes the electronic state in  a
superconductor, is a fermionic version of the two-mode squeezed state (\ref{approxstate}).
The main difference is that now 
the bosonic operators $b_{k,\sigma}^{(R/L)}$ entering in the squeezing operator in (\ref{squeezingoperator}) get replaced by fermionic counterparts. Specifically  we can write 
\begin{equation}
|\text{BCS}\rangle = \Omega |\O\rangle \label{BCSSTATE},
\end{equation}
where now 
\begin{equation} \Omega:= 
\exp\!\left[
\int d^3\bm{k}\,G_k(e^{i\theta} a_{\bm{k},\uparrow}^\dag a_{-\bm{k},\downarrow}^\dag
-e^{-i\theta} a_{-\bm{k},\downarrow} a_{\bm{k},\uparrow})
\right], \label{newomega}
\end{equation}
with $a_{\bm{k},\sigma}$ and $a_{\bm{k},\sigma}^\dag$ being the annihilation and creation operators for the electrons in the superconductor characterized by 
a spin $\sigma$ and momentum $\bm{k}$, and we have written the phase of the superconductor $\theta$ explicitly.
At the first order in the tunneling process when the electrons leak out of the superconductor, the state  $\ket{\text{BCS}}$ produces a 
correlated two-particle state whose structure reminds us of the photonic state (\ref{input}).
We wish to detect it via the Franson interferometery.

\subsection{Setting}
The setting we have in mind assumes that 
 two one-dimensional metallic leads $L$ and $R$ are attached to a superconducting source of electrons $S$, described by the BCS state (\ref{BCSSTATE}) with (\ref{newomega}), and MZIs are set up on both sides, as depicted in Fig.\ \ref{fig:Setup}.

We describe the emission of the electrons from $S$ to  $L$ and $R$ dynamically by a Hamiltonian $H=H_S+H_L+H_R+H_T$ with \cite{ref:TunnelingCohen,ref:TunnelingBardeen,ref:TunnelingPrange,ref:TunnelingAmbegaokar2,
ref:CooperBunchRapid,ref:CooperBunch,ref:CooperInt} 
\begin{gather}
H_S
=\int d^3\bm{k}
\begin{pmatrix}
a_{\bm{k},\uparrow}^\dag&
a_{-\bm{k},\downarrow}
\end{pmatrix}
\begin{pmatrix}
\medskip
\varepsilon_k&\Delta\\
\Delta^*&-\varepsilon_k
\end{pmatrix}
\begin{pmatrix}
\medskip
a_{\bm{k},\uparrow}\\
a_{-\bm{k},\downarrow}^\dag
\end{pmatrix},
\\
H_\ell=\sum_{\sigma=\uparrow,\downarrow}
\int dk\,\varepsilon_kb_{k,\sigma}^{(\ell)\dag}b_{k,\sigma}^{(\ell)}\qquad
(\ell=L,R), \label{hamell}
\displaybreak[0]\\
H_T=\sum_{\ell=L,R}\sum_{\sigma=\uparrow,\downarrow}
\int dk\int d^3\bm{k}'\,T_{k,\bm{k}'}^{(\ell)}b_{k,\sigma}^{(\ell)\dag}a_{\bm{k}',\sigma}
+\text{H.c.},
\end{gather}
where now 
 $b_{k,\sigma}^{(\ell)}$ and $b_{k,\sigma}^{(\ell)\dag}$ are the fermionic operators for the electrons propagating in lead $\ell=L,R$ before the MZIs.
$\Delta=|\Delta|e^{i\theta}$ is the gap parameter of the superconductor $S$, and 
\begin{equation}
\varepsilon_k=\frac{\hbar^2k^2}{2m}-\mu_S
\end{equation}
is the energy of an electron measured relative to the Fermi level $\mu_S$ of the superconductor $S$. 
The Hamiltonian $H_S$ can be  diagonalized by the Bogoliubov transformation
\begin{equation}
\begin{pmatrix}
\medskip
a_{\bm{k},\uparrow}\\
a_{-\bm{k},\downarrow}^\dag
\end{pmatrix}
=\begin{pmatrix}
\medskip
u_k&-v_k\\
v_k^*&u_k
\end{pmatrix}
\begin{pmatrix}
\medskip
\alpha_{\bm{k},\uparrow}\\
\alpha_{-\bm{k},\downarrow}^\dag
\end{pmatrix}
\end{equation}
with\cite{ref:Tinkham}
\begin{equation}
u_k=\frac{1}{\sqrt{2}}\sqrt{1+\frac{\varepsilon_k}{\hbar\omega_k}}
,\qquad
v_k=\frac{e^{i\theta}}{\sqrt{2}}\sqrt{1-\frac{\varepsilon_k}{\hbar\omega_k}}
,
\end{equation}
and
\begin{equation}
\hbar\omega_k=\sqrt{\varepsilon_k^2+|\Delta|^2}.
\end{equation}
The BCS state $\ket{\text{BCS}}$ can then be identified with  the vacuum state annihilated by the quasiparticle operators $\alpha_{\bm{k},\sigma}$.
Notice that the parameters $u_k$ and  $v_k$ introduced here are related to
$G_k$ of (\ref{newomega}) via the transformation 
\begin{equation}
u_k= \cos G_k,\qquad
v_k = e^{i\theta} \sin G_k.
\end{equation}
The electrons are transferred between the source and the leads by the transmission Hamiltonian $H_T$, which annihilates an electron in the superconductor $S$ and creates an electron in a lead $\ell=L,R$, and vice versa for the reverse process.
We assume that the electrons are emitted in the same manner to the left and right leads, namely, the transmission matrix elements for leads $L$ and $R$ are given, respectively, by
\begin{equation}
\begin{cases}
\medskip
T_{k,\bm{k}'}^{(L)}
=T_{k,-\bm{k}'}
e^{i(-k\bm{n}-\bm{k}')\cdot\bm{n}d/2},\\
T_{k,\bm{k}'}^{(R)}
=T_{k,\bm{k}'}
e^{-i(k\bm{n}-\bm{k}')\cdot\bm{n}d/2},
\end{cases}
\end{equation}
where $d$ is the distance between the contacts with the two leads $L$ and $R$, which are assumed to be extended in the opposite directions $-\bm{n}$ and $\bm{n}$, respectively.

The emission of the electrons starts from the initial state $\ket{\text{BCS}}_S\otimes\ket{\text{FS}}_L\otimes\ket{\text{FS}}_R$, where the electron source $S$ is in the BCS state $\ket{\text{BCS}}_S$ while the leads $L$ and $R$ are in the normal states $\ket{\text{FS}}_{L/R}$, filled with electrons up to the Fermi levels $\mu_{L/R}$, and the system eventually reaches a NESS in the long-time limit $t\to\infty$.
We look at the correlations between the detectors after the MZIs on the $L$ and $R$ arms in this NESS\@.

\subsection{Correlation functions and state}
We tune the Fermi levels of the two leads $\mu_{L/R}$ within the gap of the quasiparticle spectrum of the superconducting state, i.e., 
\begin{gather}
0<eV_L=eV_R=eV\le |\Delta|,
\\
eV_\ell=\mu_S-\mu_\ell\quad 
(\ell=L,R).
\end{gather}
In this way, the quasiparticle emission is blocked by the Fermi seas of the leads $L$ and $R$, and the Cooper-pair emission is the only allowed process.
Under this condition, collecting the contributions up to the second order in the transmission Hamiltonian $H_T$,
the only two-point correlation functions in the NESS ($t_1,t_2\to\infty$) that are nonvanishing for large $x_1,x_2$ are \cite{ref:CooperBunchRapid,ref:CooperBunch,ref:CooperInt}
\begin{widetext}
\begin{align}
&\langle
\psi_\uparrow^{(\ell_1)}(x_1,t_1)
\psi_\downarrow^{(\ell_2)}(x_2,t_2)
\rangle
=-\langle
\psi_\downarrow^{(\ell_1)}(x_1,t_1)
\psi_\uparrow^{(\ell_2)}(x_2,t_2)
\rangle
\nonumber\displaybreak[0]\\
&\qquad\qquad\qquad%
\simeq -\frac{im^2}{\hbar^4}\int_{-eV}^{eV}dE\int d^3\bm{k}\,
u_kv_k
\frac{
T_{\bar{k}(E),-\bm{k}}^{(\ell_1)}
T_{\bar{k}(-E),\bm{k}}^{(\ell_2)}
}{\bar{k}(E)\bar{k}(-E)}
\frac{2\hbar\omega_k}{E^2-\hbar^2\omega_k^2+i0^+}
e^{i[\bar{k}(E)x_1+\bar{k}(-E)x_2]}
e^{-iE(t_1-t_2)/\hbar}
\label{eqn:Chi0}
\end{align}
\end{widetext}
and their complex conjugates, where $
\psi_\sigma^{(\ell)}(x,t)
$
is the field operator for the electrons before the MZI on arm $\ell\,(=L,R)$ in the Heisenberg representation, 
\begin{gather}
\psi_\sigma^{(\ell)}(x,t)
=e^{iHt/\hbar}\psi_\sigma^{(\ell)}(x)e^{-iHt/\hbar},
\displaybreak[0]\\
\psi_\sigma^{(\ell)}(x)
=\int\frac{dk}{\sqrt{2\pi}}b_{k,\sigma}^{(\ell)}e^{ikx},
\end{gather}
which is the counterpart of (\ref{free}) but with $b_{p,\sigma}^{(\ell)}$ being now the fermionic operator of (\ref{hamell}), 
and  
$
\bar{k}(E)=k_F\sqrt{1+E/\mu_S}
$ with the Fermi wave vector
$
k_F=2\pi/\lambda_F=\sqrt{2m\mu_S/\hbar^2}
$.
Under the assumptions 
$
k_Fd\gg1
$ and 
$
|\Delta|/\mu_S\ll1
$, we get 
\begin{align}
&\langle
\psi_\uparrow^{(\ell_1)}(x_1,t_1)
\psi_\downarrow^{(\ell_2)}(x_2,t_2)
\rangle
\nonumber\\
&\quad
\simeq\frac{1}{\sqrt{2\pi}}e^{ik_F(x_1+x_2-d)}\Phi^{(\ell_1,\ell_2)}\bm{(}(x_1-x_2)-v_F(t_1-t_2)\bm{)},
\end{align}
where
\begin{align}
\Phi(x)
&=\Phi^{(L,R)}(x)=\Phi^{(R,L)}(x)\nonumber\\
&=\frac{A}{2\pi\xi}
\int_{-eV/|\Delta|}^{eV/|\Delta|}
\frac{ds}{\sqrt{1-s^2}}
e^{-(d/\pi\xi)\sqrt{1-s^2}}e^{isx/\pi\xi},
\label{eqn:ChiInt}
\displaybreak[0]\\
\tilde{\Phi}(x)
&=\Phi^{(L,L)}(x)=\Phi^{(R,R)}(x)\nonumber\\
&=\frac{\tilde{A}}{2\pi\xi}
\int_{-eV/|\Delta|}^{eV/|\Delta|}
\frac{ds}{\sqrt{1-s^2}}
e^{isx/\pi\xi},
\label{eqn:ChiIntTilde}
\end{align}
with dimensionless constants
$A=e^{i\theta}(2\pi)^{5/2}(m^2/\hbar^4)$ ${}\times(T_{k_F,k_F\bm{n}}^2e^{ik_Fd}-T_{k_F,-k_F\bm{n}}^2e^{-ik_Fd})/k_Fd$
and
$\tilde{A}=i(2\pi)^{3/2}(m^2/\hbar^4)\int d^2\hat{\bm{n}}\,T_{k_F,k_F\bm{n}}T_{k_F,-k_F\bm{n}}$, and Pippard's length
\begin{equation}
\xi=\frac{\hbar v_F}{\pi|\Delta|}=\frac{2\mu_S}{\pi k_F|\Delta|},
\end{equation}
characterizing the extension of a Cooper pair,\cite{ref:Tinkham} with $v_F=\hbar k_F/m$ being the Fermi velocity.
In the following, we assume $d/\pi\xi\ll1$, which is typically the case: see Ref.~\onlinecite{ref:CooperSplitterNature} and the caption of Fig.\ \ref{fig:chi2}.
Then, in the zero-bias regime $eV\ll|\Delta|$,
the pair wave functions behave as
\begin{equation}
\Phi(x)/A
\simeq
\tilde{\Phi}(x)/\tilde{A}
\simeq
\frac{eV}{\pi\xi|\Delta|}\sinc[(x/\pi\xi)(eV/|\Delta|)]
\label{eqn:ChiAna}
\end{equation}
for $x/\pi\xi\gg1$, which as a function of $x$ has a width $\Delta x \simeq \pi^2 \xi |\Delta |/eV=\pi\hbar v_F/eV$.  
For $eV=|\Delta|$, on the other hand, they are estimated to be
\begin{equation}
\Phi(x)/A
\simeq
\tilde{\Phi}(x)/\tilde{A}
\simeq
\frac{1}{2\xi}J_0(x/\pi\xi)
\label{eqn:ChiBessel}
\end{equation}
for the whole range of $x$, where $J_0(x)$ is a Bessel function.
This has a width $\Delta x\simeq2.4\,\pi\xi$.
See Fig.\ \ref{fig:chi2}, where $\Phi(x)$ is shown as a function of $x$ and $eV$.
%%%%%%%%%%%%%%%%%%%%%%%%%%%%%%%%%
\begin{figure}[t]
\includegraphics{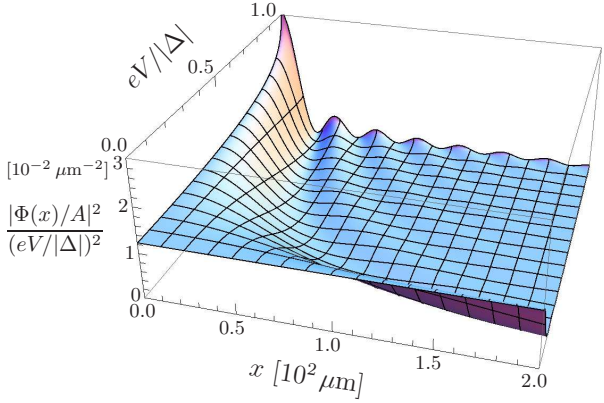}
\caption{(Color online) The pair wave function $|\Phi(x)|^2$ [normalized by $(AeV/|\Delta|)^2$] as a function of $x$ and $eV$, on the basis of the numerical integration of (\ref{eqn:ChiInt}). 
The parameters are $|\Delta|=150\,\mu\text{eV}$, $\lambda_F=2\pi/k_F=0.36\,\text{nm}$, $\mu_S=11.6\,\text{eV}$, $d=150\,\text{nm}$, and $\xi=2.8\,\mu\text{m}$ for an Al superconductor.\cite{ref:CooperSplitterNature}}
\label{fig:chi2}
\end{figure}
%%%%%%%%%%%%%%%%%%%%%%%%%%%%%%%%%

Notice that $\langle\psi_\sigma^{(\ell)}(x,t)\rangle=0$,\cite{ref:CooperBunchRapid,ref:CooperBunch,ref:CooperInt} and recall that the correlation functions in (\ref{eqn:Chi0}) and their complex conjugates are the only nonvanishing two-point correlation functions up to the second order in the transmission Hamiltonian $H_T$.
This implies that the state of the electrons in the leads $L$ and $R$ emitted from the superconductor $S$ is expressed, up to the second order in $H_T$, as (\ref{approxstate}) with\cite{ref:Samuelsson-SuperBell}
\begin{equation}
\ket{\Psi^{(2)}}
:=
\sum_{\ell_1,\ell_2}
 \int dk \,\Psi_k^{(\ell_1,\ell_2)}b_{k_F+k,\downarrow}^{(\ell_1)\dag}b_{k_F-k,\uparrow}^{(\ell_2)\dag}\ket{\O},
\label{inputS}
\end{equation}
where $\Psi_k^{(\ell_1,\ell_2)}$ are the Fourier transforms of $\Phi^{(\ell_1,\ell_2)}(x)$ as in (\ref{eqn:PhiPsi}).
This should be compared with (\ref{input}) for the optical case, where the polarization state of a pair of photons is symmetric, while in the superconductor case in (\ref{inputS}) the spin state of a pair of electrons is antisymmetric (the singlet state). On the other hand, the states of their spatial degrees of freedom are symmetric in both cases.

\subsection{Coincidences}
We can now investigate the Franson interferometry for the electrons emitted from a superconductor.
Using (\ref{freeout}) the coincidence rate between detectors $(L,-)$ and $(R,-)$ is found to be ruled by an expression which reminds us of (\ref{eqn:I}) for the optical scenario, i.e., 
\begin{align}
I^{(2)}(\tau)
 = \frac{1}{16\pi}
|& 2 \Phi(v_F\tau)\cos\phi
-e^{i\varphi}\Phi(v_F\tau+\delta L) 
\nonumber\\
&\qquad\qquad\quad
{}-e^{-i\varphi}\Phi(v_F\tau-\delta L)|^2, 
\end{align}
where now
\begin{equation}
\phi:=k_F\delta L+(\varphi_L+\varphi_R)/2
\end{equation}
instead of (\ref{varphi}).
The phases $\varphi_{L/R}$ can be induced by the Aharonov-Bohm effect, by applying magnetic field perpendicular to the MZIs.\cite{ref:Antibunching-Ji,ref:MachZenderElectron-PRL100,ref:MachZenderElectron-PRB78,ref:MachZenderElectron-PRB79,ref:Antibunching-Neder,ref:BuettikerReview2000,ref:Samuelsson-HBT,ref:SamuelssonButtikerReview}
It is clear that the formulas for the optical case are translated to those for the electric case by simply substituting $k_0\to k_F$ and $c\to v_F$, and the pair wave function $\Phi(x)$ of the emitted Cooper pair is given by (\ref{eqn:ChiInt}).
The single-electron detection rate at $(L/R,-)$, on the other hand, is proportional to
\begin{align}
&I^{(1)}(x_{L/R},t)
\nonumber\\
&\quad
=\frac{\Lambda}{4\pi}\int dx\,|\Phi(x)-e^{i(\varphi_{L/R}+k_F\delta L)}\Phi(x+\delta L)|^2
\end{align}
with $\Lambda=1+|\tilde{A}/A|^2$, which is to be compared with (\ref{eqn:I1}) for the optical scenario.
The correction proportional to $|\tilde{A}/A|^2$ is due to the possible emission of a Cooper pair into the same lead, either $L$ or $R$, which was assumed to be absent when we considered the optical scenario.
Such processes would be suppressed by employing the Cooper-pair splitter,\cite{ref:CooperSplitterPRB-Loss,ref:BurkardReview-JPC2007} realized in Refs.~\onlinecite{ref:CooperSplitterNature} and~\onlinecite{ref:CooperSplitterPRL}.

It is now clear that the formulas for the coincidence rate listed in Sec.\ \ref{sec:one} can be applied to the present electric scenario by the small changes $k_0\to k_F$ and $c\to v_F$, and the introduction of the factor $\Lambda$.
Then, by employing the strategies outlined in Secs.\ \ref{sec:FineReso}--\ref{sec:random}, we can probe the coherence properties of the emitted Cooper pair.
That is:
\begin{enumerate}
\item We perform the correlation measurements, counting the coincidences between output ports $(L,-)$ and $(R,-)$ of the Franson interferometer, in the \textit{low imbalance regime} $\delta L\ll\Delta x$, with $\varphi_L=\varphi_R$.
In this case, the average number of coincidences $N^{(2)}$ oscillates as a function of $\phi$ with period $2\pi$.
See (\ref{low}) for fine time resolution $\Delta \tau\ll\Delta x/c$ and (\ref{ratedeflong1}) for low time resolution $\Delta \tau\gg\Delta x/c$.
This oscillation with period $2\pi$ is due to the \textit{single}-electron interference through each MZI\@.

\item We then move to the \textit{high imbalance regime} $\delta L\gg\Delta x$, with $\varphi_L=\varphi_R$, and observe that $N^{(2)}$ oscillates as a function of $\phi$ with period $\pi$.
See (\ref{high}) for fine time resolution $\Delta \tau\ll\Delta x/c$ and (\ref{ratedeflong2}) for low time resolution $\Delta \tau\gg\Delta x/c$.
This oscillation with period $\pi$ in turn is due to the interference between the LL and SS processes  [Fig.\ \ref{fig:InterferingProcs}(a)], and it can be regarded as the MZ interference of a Cooper pair \textit{as a single object}, as if a Cooper pair as a whole goes through the long and short paths of a MZI and interferes, even though it is actually split into two different branches and its constituent electrons go through different MZIs.
From the interference fringes in $N^{(2)}$ as a function of $\delta L$, we can measure the de Broglie wavelength of the center-of-mass motion of the emitted Cooper pair, $\lambda_\text{pair}=2\pi\hbar/p_\text{pair}=\pi/k_F=\lambda_F/2$, i.e., the de Broglie wavelength of the Cooper pair as a single (composite) particle.

\item In addition, by observing the transition between the low and high imbalance regimes, we can measure the pair correlation length $\Delta x$ of the emitted Cooper pair.
Since $\Delta x$ is proportional to Pippard's length $\xi$ [see (\ref{eqn:ChiInt}), (\ref{eqn:ChiAna}), and (\ref{eqn:ChiBessel})], we can also estimate Pippard's length $\xi$  characterizing the size of a Cooper pair in the superconducting source. 
\end{enumerate}
In these ways, we can probe the coherences of both the internal and center-of-mass degrees of freedom of the emitted Cooper pair.

It could be, however, a bit challenging to enforce the high imbalance condition $\delta L\gg\Delta x$ in practice.
The latter would be typically of the order of $\Delta x\sim10\,\mu\text{m}$, whereas the wavelengths of the electrons passing through the MZIs would be $\lambda_F\sim0.5\,\text{nm}$ (see Fig.\ \ref{fig:chi2} and its caption).
It would be difficult to take such a large path difference $\delta L$ for each MZI\@.
Even in such a case, we can employ the randomization procedure outlined in Sec.\ \ref{sec:random}:
\begin{enumerate}
\item[(iv)] We perform the correlation measurements in the low imbalance regime $\delta L\ll\Delta x$, with $\varphi_L+\varphi_R$ kept constant but $\varphi=(\varphi_L-\varphi_R)/2$ randomized.
Then, $N^{(2)}$ behaves as (\ref{eqn:N2Randomized}) when the time resolution is fine, $\Delta\tau\ll\Delta x/c$, while it behaves as (\ref{ratedeflong2}) when the time resolution is low, $\Delta\tau\gg\Delta x/c$.
In both cases, $N^{(2)}$ oscillates as a function of $\phi$ with period $\pi$ (with half visibility), and we can probe the de Broglie wavelength of the emitted Cooper pair. 
The estimation of Pippard's length $\xi$ is problematic, but it is at least in principle possible by studying the variation of the visibility of the oscillations of $N^{(2)}$ as a function of the coincidence time  window $\Delta \tau$.
Indeed from (\ref{ratedeflong2})  and  (\ref{eqn:N2Randomized}) we notice  that, while for  $\Delta \tau \ll \Delta x/c$ the visibility of $N^{(2)}$ (or equivalently the visibility of the ratio ${\cal R}$) is increasing with $\Delta \tau$, 
for $\Delta \tau \gg \Delta x/c$ it saturates to a constant value.
Monitoring this change of behavior enables one to estimate $\Delta x$, i.e., Pippard's length $\xi$.
\end{enumerate}

As for the spin degrees of freedom of the Cooper pair, the coincidence ratio $\mathcal{R}$ defined in (\ref{correlations}) 
exceeds unity, for $\phi=0$, in the ideal situation where the time resolutions of the detectors are fine and the accumulation time duration $T$ is short.
This is bunching and is due to the singlet spin correlation of the Cooper pair.
Therefore, by observing the bunching, we can detect the spin entanglement of the Cooper pair.\cite{ref:BuettikerReview2000,PhysRevB.61.R16303,PhysRevB.65.075317,PhysRevLett.88.197001,PhysRevB.66.161320,PhysRevB.69.125326,PhysRevB.70.115330,ref:BurkardReview-JPC2007,ref:CooperBunchRapid,ref:CooperBunch,ref:CooperInt}
For this purpose, however, the Franson setting is not essential, and in addition the bunching is lost if the time resolutions of the detectors are not very fine and the accumulation time duration $T$ is relatively long, as considered in the present paper.

\section{Conclusions}\label{Conclusions}
In this paper we reviewed some of the features of the Franson interferometric scheme,\cite{ref:FransonInt} discussing how it can be used to discriminate the two-particle coherence effects from the single-particle coherence effects in different experimental settings. Subsequently, exploiting a formal equivalence between 
the state of a  Cooper pair emitted via tunneling from a superconductor into external metallic contacts and the two-mode squeezed state generated in a parametric process, 
we have analyzed an application of this setup for electronic sources.

As already mentioned 
the emission of the electrons comprising a Cooper pair into two separate metallic leads has been realized recently.\cite{ref:CooperSplitterNature,ref:CooperSplitterPRL}
The main technical challenge for the implementation of the proposed scheme 
is posed by the realization of the  beam splitter transformations and by the ability of controlling the relative phase accumulated in the propagation of the electrons through the MZIs.  
A beam splitter for electrons was realized in Refs.~\onlinecite{ref:Antibunching-Yamamoto} and \onlinecite{ref:Antibunching-Oliver} in a GaAs high-mobility two-dimensional electron gas system, where two input and two output ports are defined by applying negative bias voltages to gates that deplete underlying electrons, and the splitting ratio of the beam splitter is adjusted by applying an appropriate small bias voltage to another thin gate.
A correlation experiment, which is an important ingredient for the Franson interferometry, was carried out in such a system.\cite{ref:Antibunching-Yamamoto,ref:Antibunching-Oliver} However, for this type of beam splitter, the back reflection of electrons to the input ports is inevitable, resulting in a reduction of the number of coincidence events. 
This problem is absent in the beam splitter realized for the chiral edge current of a two-dimensional electron gas in the quantum Hall effect regime.\cite{ref:Antibunching-Ji,ref:MachZenderElectron-PRL100,ref:MachZenderElectron-PRB78,ref:MachZenderElectron-PRB79,ref:Antibunching-Henny,ref:Antibunching-Neder}
The quasi-one-dimensionality of the edge current is also appropriate for constructing interferometers, and MZIs were actually realized\cite{ref:Antibunching-Ji,ref:MachZenderElectron-PRL100,ref:MachZenderElectron-PRB78,ref:MachZenderElectron-PRB79} and correlation experiments were also performed\cite{ref:Antibunching-Henny,ref:Antibunching-Neder} in such systems. Furthermore the possibility of interfacing quantum Hall bars with superconductors at high magnetic fields has been recently established in Refs.~\onlinecite{PhysRevB.62.7477,PhysRevLett.95.107001,springerlink:10.1007/s00339-007-4192-5,PhysRevB.76.115313}. 
Using chiral edge currents instead of metallic leads appears hence to  be a possible option for the implementation of our setup.

\acknowledgments
We thank Rosario Fazio for useful comments and discussions. 
This work was supported by MIUR through FIRB-IDEAS Project No.\ RBID08B3FM and by the Joint Italian-Japanese Laboratory on ``Quantum Technologies: Information, Communication and Computation'' of the Italian Ministry of Foreign Affairs.
KY was supported by the Program to Disseminate Tenure Tracking System and the Grant-in-Aid for Young Scientists (B) (No.\ 21740294) both from the Ministry of Education, Culture, Sports, Science and Technology, Japan.

%\bibliography{FransonBCS}

\begin{thebibliography}{10}

\bibitem{ref:FransonReview2}
J.~D. Franson, in \textit{Proceedings of the Ninth Rochester Conference on
  Coherence and Quantum Optics}, edited by N.~P. Bigelow, J.~H. Eberly, and
  C.~R. {Stroud, Jr.} (Optical Society of America, Washington, DC, 2008), pp.\
  178--190; arXiv:0707.0475 [quant-ph].

\bibitem{TITTEL}
W. Tittel and G. Weihs, Quant. Inf. Comput. \textbf{1},  3  (2001).

\bibitem{ref:FransonInt}
J.~D. Franson, Phys. Rev. Lett. \textbf{62},  2205  (1989).

\bibitem{ref:Jacobson-PRL1995}
J. Jacobson, G. Bj\"ork, I. Chuang, and Y. Yamamoto, Phys. Rev. Lett.
  \textbf{74},  4835  (1995).

\bibitem{ref:Barnett1998}
S.~M. Barnett, N. Imoto, and B. Huttner, J. Mod. Opt. \textbf{45},  2217
  (1998).

\bibitem{ref:Fonseca-PRL1999}
E.~J.~S. Fonseca, C.~H. Monken, and S. P\'adua, Phys. Rev. Lett. \textbf{82},
  2868  (1999).

\bibitem{ref:Edamatsu-PRL2002}
K. Edamatsu, R. Shimizu, and T. Itoh, Phys. Rev. Lett. \textbf{89},  213601
  (2002).

\bibitem{EXP}
P.~G. Kwiat, W.~A. Vareka, C.~K. Hong, H. Nathel, and R.~Y. Chiao, Phys. Rev. A
  \textbf{41},  R2910  (1990).

\bibitem{PhysRevLett.65.321}
Z.~Y. Ou, X.~Y. Zou, L.~J. Wang, and L. Mandel, Phys. Rev. Lett. \textbf{65},
  321  (1990).

\bibitem{PhysRevLett.66.1142}
J. Brendel, E. Mohler, and W. Martienssen, Phys. Rev. Lett. \textbf{66},  1142
  (1991).

\bibitem{PhysRevA.44.4552}
J.~D. Franson, Phys. Rev. A \textbf{44},  4552  (1991).

\bibitem{PhysRevA.45.2052}
J.~G. Rarity and P.~R. Tapster, Phys. Rev. A \textbf{45},  2052  (1992).

\bibitem{ref:Shih-PRA1993}
Y.~H. Shih, A.~V. Sergienko, and M.~H. Rubin, Phys. Rev. A \textbf{47},  1288
  (1993).

\bibitem{PhysRevA.47.R2472}
P.~G. Kwiat, A.~M. Steinberg, and R.~Y. Chiao, Phys. Rev. A \textbf{47},  R2472
   (1993).

\bibitem{PhysRevA.77.021801}
A.~K. Jha, M.~N. O'Sullivan, K.~W.~C. Chan, and R.~W. Boyd, Phys. Rev. A
  \textbf{77},  021801(R)  (2008).

\bibitem{PhysRevLett.96.090402}
J.~D. Franson, Phys. Rev. Lett. \textbf{96},  090402  (2006).

\bibitem{ref:Plasmons}
S. Fasel, M. Halder, N. Gisin, and H. Zbinden, New J. Phys. \textbf{8},  13
  (2006).

\bibitem{PhysRevLett.92.167901}
V. Scarani, N. Gisin, and S. Popescu, Phys. Rev. Lett. \textbf{92},  167901
  (2004).

\bibitem{PhysRevLett.103.076804}
J. Splettstoesser, M. Moskalets, and M. B\"uttiker, Phys. Rev. Lett.
  \textbf{103},  076804  (2009).

\bibitem{ref:LucaVittorioSaro-TimeBin}
L. Chirolli, V. Giovannetti, R. Fazio, and V. Scarani, Phys. Rev. B
  \textbf{84},  195307  (2011).

\bibitem{ref:Antibunching-Ji}
Y. Ji, Y. Chung, D. Sprinzak, M. Heiblum, D. Mahalu, and H. Shtrikman, Nature
  (London) \textbf{422},  415  (2003).

\bibitem{ref:MachZenderElectron-PRL100}
P. Roulleau, F. Portier, P. Roche, A. Cavanna, G. Faini, U. Gennser, and D.
  Mailly, Phys. Rev. Lett. \textbf{100},  126802  (2008).

\bibitem{ref:MachZenderElectron-PRB78}
L.~V. Litvin, A. Helzel, H.-P. Tranitz, W. Wegscheider, and C. Strunk, Phys.
  Rev. B \textbf{78},  075303  (2008).

\bibitem{ref:MachZenderElectron-PRB79}
E. Bieri, M. Weiss, O. G\"oktas, M. Hauser, C. Sch\"onenberger, and S.
  Oberholzer, Phys. Rev. B \textbf{79},  245324  (2009).

\bibitem{ref:Antibunching-Yamamoto}
R.~C. Liu, B. Odom, Y. Yamamoto, and S. Tarucha, Nature (London) \textbf{391},
  263  (1998).

\bibitem{ref:Antibunching-Oliver}
W.~D. Oliver, J. Kim, R.~C. Liu, and Y. Yamamoto, Science \textbf{284},  299
  (1999).

\bibitem{ref:Antibunching-Henny}
M. Henny, S. Oberholzer, C. Strunk, T. Heinzel, K. Ensslin, M. Holland, and C.
  Sch\"onenberger, Science \textbf{284},  296  (1999).

\bibitem{ref:Antibunching-Neder}
I. Neder, N. Ofek, Y. Chung, M. Heiblum, D. Mahalu, and V. Umansky, Nature
  (London) \textbf{448},  333  (2007).

\bibitem{ref:BuettikerReview2000}
{Ya. M. Blanter} and M. B\"uttiker, Phys. Rep. \textbf{336},  1  (2000).

\bibitem{ref:Samuelsson-HBT}
P. Samuelsson, E.~V. Sukhorukov, and M. B\"uttiker, Phys. Rev. Lett.
  \textbf{92},  026805  (2004).

\bibitem{ref:SamuelssonButtikerReview}
J. Splettstoesser, P. Samuelsson, M. Moskalets, and M. B\"uttiker, J. Phys. A
  \textbf{43},  354027  (2010).

\bibitem{ref:CooperSplitterNature}
L. Hofstetter, S. Csonka, J. Nyg\r{a}rd, and C. Sch\"onenberger, Nature
  (London) \textbf{461},  960  (2009).

\bibitem{ref:CooperSplitterPRL}
L.~G. Herrmann, F. Portier, P. Roche, A.~L. Yeyati, T. Kontos, and C. Strunk,
  Phys. Rev. Lett. \textbf{104},  026801  (2010).


\bibitem{PhysRevB.61.R16303}
G. Burkard, D. Loss, and E.~V. Sukhorukov, Phys. Rev. B \textbf{61},  R16303
  (2000).

\bibitem{PhysRevB.65.075317}
F. Taddei and R. Fazio, Phys. Rev. B \textbf{65},  075317  (2002).

\bibitem{PhysRevLett.88.197001}
J. B\"orlin, W. Belzig, and C. Bruder, Phys. Rev. Lett. \textbf{88},  197001
  (2002).

\bibitem{PhysRevB.66.161320}
N.~M. Chtchelkatchev, G. Blatter, G.~B. Lesovik, and T. Martin, Phys. Rev. B
  \textbf{66},  161320(R)  (2002).

\bibitem{PhysRevB.69.125326}
L. Faoro, F. Taddei, and R. Fazio, Phys. Rev. B \textbf{69},  125326  (2004).

\bibitem{PhysRevB.70.115330}
P. Samuelsson, E.~V. Sukhorukov, and M. B\"uttiker, Phys. Rev. B \textbf{70},
  115330  (2004).

\bibitem{ref:BurkardReview-JPC2007}
G. Burkard, J. Phys.: Condens. Matter \textbf{19},  233202  (2007).

\bibitem{ref:CooperBunchRapid}
K. Yuasa, P. Facchi, R. Fazio, H. Nakazato, I. Ohba, S. Pascazio, and S.
  Tasaki, Phys. Rev. B \textbf{79},  180503(R)  (2009).

\bibitem{ref:CooperBunch}
K. Yuasa, Phys. Rev. B \textbf{80},  104516  (2009).

\bibitem{ref:CooperInt}
M. Iazzi and K. Yuasa, Phys. Rev. B \textbf{81},  172501  (2010).

\bibitem{ref:KwiatPDCEntanglement}
P.~G. Kwiat, K. Mattle, H. Weinfurter, A. Zeilinger, A.~V. Sergienko, and Y.
  Shih, Phys. Rev. Lett. \textbf{75},  4337  (1995).

\bibitem{ref:Grice-PRA1997}
W.~P. Grice and I.~A. Walmsley, Phys. Rev. A \textbf{56},  1627  (1997).

\bibitem{NOTE}
In the optical domain this is a rather challenging requirement as it requires
  operating electronic devices with sub-pico second response times (see, e.g.,
  Refs.~\onlinecite{EXP,PhysRevLett.65.321,PhysRevLett.66.1142,PhysRevA.44.4552,PhysRevA.45.2052,ref:Shih-PRA1993,PhysRevA.47.R2472,PhysRevA.77.021801}).
  On the other hand it provides a clean analytical result which helps clarify
  the working principle of the Franson setting.

\bibitem{NOTE1}
It is worth pointing out that analogous results hold even if $\Delta
  \tau\gg\Delta x/c$, as long as the imbalance length $\delta L$ satisfies the
  condition $\Delta \tau\ll \delta L/c$: in such a limit Eq.\ (\ref{high})
  still applies with $|{\Phi}(0)|^2\Delta \tau$ being replaced by the quantity
  $\mathcal{X}(0)/c$.

\bibitem{ref:Tinkham}
M. Tinkham, \textit{Introduction to Superconductivity}, 2nd ed. (Dover
  Publications, New York, 1996).

\bibitem{ref:TunnelingCohen}
M.~H. Cohen, L.~M. Falicov, and J.~C. Phillips, Phys. Rev. Lett. \textbf{8},
  316  (1962).

\bibitem{ref:TunnelingBardeen}
J. Bardeen, Phys. Rev. Lett. \textbf{9},  147  (1962).

\bibitem{ref:TunnelingPrange}
R.~E. Prange, Phys. Rev. \textbf{131},  1083  (1963).

\bibitem{ref:TunnelingAmbegaokar2}
V. Ambegaokar and A. Baratoff, Phys. Rev. Lett. \textbf{10}, 486 (1963);
  \textbf{11}, 104(E) (1963).

\bibitem{ref:Samuelsson-SuperBell}
P. Samuelsson, E.~V. Sukhorukov, and M. B\"uttiker, Phys. Rev. Lett.
  \textbf{91},  157002  (2003).

\bibitem{ref:CooperSplitterPRB-Loss}
P. Recher, E.~V. Sukhorukov, and D. Loss, Phys. Rev. B \textbf{63},  165314
  (2001).

\bibitem{PhysRevB.62.7477}
Y. Asano and T. Yuito, Phys. Rev. B \textbf{62},  7477  (2000).

\bibitem{PhysRevLett.95.107001}
J. Eroms, D. Weiss, J. De Boeck, G. Borghs, and U. Z\"ulicke, Phys. Rev. Lett.
  \textbf{95},  107001  (2005).

\bibitem{springerlink:10.1007/s00339-007-4192-5}
J. Eroms and D. Weiss, Appl. Phys. A \textbf{89},  639  (2007).

\bibitem{PhysRevB.76.115313}
I.~E. Batov, T. Sch\"apers, N.~M. Chtchelkatchev, H. Hardtdegen, and A.~V.
  Ustinov, Phys. Rev. B \textbf{76},  115313  (2007).

\end{thebibliography}
%\end{document}

\end{document}